\documentstyle[times,pramana,epsfig,floats]{ias}

\def\PRC{{\em Phys. Rev.} {\bf C}}
\def\NPA{{\em Nucl. Phys.} {\bf A}}
\def\PLB{{\em Phys. Lett.} {\bf B}}
\def\PRL{{\em Phys. Rev. Lett.} }

\newcommand{\be}{\begin{equation}}
\newcommand{\ee}{\end{equation}}
\newcommand{\bea}{\begin{eqnarray}}
\newcommand{\eea}{\end{eqnarray}}

\begin{document}
\title{ Hadrons in Medium}
\author{U. Mosel}
\address{
Institut fuer Theoretische Physik, Universitaet Giessen\\
D-35392 Giessen, Germany} \maketitle \abstract{ In these lectures I
first give a motivation for investigations of in-medium properties
of hadrons. I discuss the relevant symmetries of QCD and how they
might affect the observed hadron properties. I then discuss at
length the observable consequences of in-medium changes of hadronic
properties in reactions with elementary probes, and in particular
photons, on nuclei. Here I put an emphasis on new experiments on
changes of the $\sigma$ and $\omega$ mesons in medium.}


\section{Introduction}

Studies of in-medium properties of hadrons are driven by a number of
- partly connected - motivations. A first motivation for the study
of hadronic in-medium properties is provided by our interest in
understanding the structure of large dense systems, such as the
interior of stars. This structure obviously depends on the
composition of stellar matter and its interactions (for a recent
review see \cite{Weber}).

The second motivation is based on the expectation that changes of
hadronic properties in medium can be precursor phenomena to changes
in the phase structure of nuclear matter. Here the transition to the
chirally symmetric phase, that exhibits manifestly the symmetries of
the underlying theory of strong interactions, i.e. QCD, is of
particular interest. Present day's ultrarelativic heavy-ion
collisions explore this phase boundary in the limit of high
temperatures ($T \approx$ 170 MeV) and low densities. The other
limit (low temperatures and high densities) is harder to reach,
although the older AGS heavy-ion experiments and the planned CBM
experiment at the new FAIR facility \cite{FAIR} may yield insight
into this area. However, even in these experiments the temperatures
reached are still sizeable ($T \approx $ 120 MeV). At very low
temperatures the only feasible method seems to be the exploration of
the hadronic structure inside ordinary nuclei, at the prize of a low
density. Here the temperature is $T=0$ and the density at most
equals the equilibrium density of nuclear matter, $\rho_0$. It is
thus of great interest to explore if such low densities can already
give precursor signals for chiral symmetry restoration.

That hadrons can indeed change their properties and couplings in the
nuclear medium has been well known to nuclear physicists since the
days of the Delta-hole model that dealt with the changes of the
properties of the pion and Delta-resonance inside nuclei
\cite{Ericsson-Weise}. Due to the predominant $p$-wave interaction
of pions with nucleons one observes here a lowering of the pion
branch with increasing pion-momentum and nucleon-density. This
effect can be seen in optical model analyses of pion scattering on
nuclei, but the absorptive part of the $\pi$-nucleus interaction
limits the sensitivity to small densities. More recently,
experiments at the FSR at GSI have shown that also the mass of a
pion at rest in the nuclear medium differs from its value in vacuum
\cite{Kienle}. This is interesting since there are also recent
experiments \cite{TAPSsigma} that look for in-medium changes of the
$\sigma$ meson, the chiral partner of the pion. Any comparison of
scalar and pseudoscalar strength could thus give information about
the degree of chiral symmetry restoration in nuclear matter.

In addition, experiments for charged kaon production at GSI
\cite{KAOS} have given some evidence for the theoretically predicted
lowering of the $K^-$ mass in medium and the (weaker) rising of the
$K^+$ mass. State-of-the-art calculations of the in-medium
properties of kaons have shown that the usual quasi-particle
approximation for these particles is no longer justified inside
nuclear matter where they acquire a broad spectral function
\cite{Lutz,Tolos}.

At higher energies, at the CERN SPS and most recently at the
Brookhaven RHIC, in-medium changes of vector mesons have found
increased interest. This was due to two main reasons. First, these
mesons are the first excitations of the QCD vacuum that are not
protected by chiral symmetry, as the pions as Goldstone bosons are,
and should therefore be more directly related to the chiral
condensates. This expectation was triggered by the original work of
Hatsuda and Lee \cite{Hatsuda} that, based on QCD sum rules,
predicted a significant lowering of the vector meson masses with
increasing density, the effect being as large as 30\% already at
nuclear matter equilibrium density. As discussed in
\cite{Leupold,LeupoldMosel} this original prediction strongly
depended on simplifying assumptions for the spectral function of the
particles involved. When more realistic spectral shapes are used the
QCD sum rule gives only certain restrictions on mass and width of
the particles involved, but does not fix the latter; for that
hadronic models are needed. In particular, for the $\rho$ meson it
turned out that the broadening is more dominant than a mass-shift
\cite{Postneu}.

The second reason for the interest in vector meson production stems
from the fact that these mesons couple strongly to the photon so
that electromagnetic signals can yield information about properties
of hadrons deeply embedded into nuclear matter. Indeed, the CERES
experiment \cite{CERES} has found a considerable excess of dileptons
in an invariant mass range from $\approx 300$ MeV to $\approx 700$
MeV as compared to expectations based on the assumption of freely
radiating mesons.

This result has found an explanation in terms of a shift of the
$\rho$ meson spectral function down to lower masses, as expected
from theory (see, e.g., \cite{Postneu,Peters,Post,Wambach}).
However, the actual reason for the observed dilepton excess is far
from clear. Both models that just shift the pole mass of the vector
meson as well as those that also modify the spectral shape have
successfully explained the data \cite{Cassingdil,RappWam,Rapp}; in
addition, even a calculation that just used the free radiation rates
with their -- often quite large -- experimental uncertainties was
compatible with the observations \cite{Koch}. There are also
calculations that attribute the observed effect to radiation from a
quark-gluon plasma \cite{Renk}.

The more recent experimental results on a change of the rho-meson
properties in-medium obtained in an ultrarelativistic $Au + Au$
collision by the STAR collaboration at RHIC \cite{STAR} show a
downward shift of the $\rho$-meson pole mass by about 70 MeV. Since
also the $p + p$ data obtained in the same experiment show a
similar, though slightly less pronounced (-40 MeV), shift of the
$\rho$-meson mass, phase-space distortions of the $\rho$-meson
spectral shape may be at least partly responsible for the observed
mass shift. In the heavy-ion experiment then a number of additional
effects, mainly by dynamical interactions with surrounding matter,
may contribute, but are hard to separate from the more mundane phase
space effects \cite{Shuryak-Brown}.

For  both experiments quite different model calculations tend to
explain the data, though often with some model assumptions. Their
theoretical input is sufficiently different as to make the inverse
conclusion that the data prove one or another of the proposed
explanations impossible. In particular, if one is after some
`exotic` effect like a signal for the QGP, one has to make sure that
one understands the `classical` contributions to the observed
signals very well.

I have therefore already some years ago proposed to look for the
theoretically predicted changes of vector meson properties inside
the nuclear medium in reactions on normal nuclei with more
microscopic probes \cite{Hirschegg97,Hirschegg}. Of course, the
nuclear density felt by the vector mesons in such experiments lies
much below the equilibrium density of nuclear matter, $\rho_0$, so
that naively any density-dependent effects are expected to be much
smaller than in heavy-ion reactions.

On the other hand, there is a big advantage to these experiments:
they proceed with the spectator matter being close to its
equilibrium state. This is essential because all theoretical
predictions of in-medium properties of hadrons are based on an
equilibrium model in which the hadron (vector meson) under
investigation is embedded in cold nuclear matter in equilibrium and
with infinite extension. However, a relativistic heavy-ion reaction
proceeds -- at least initially -- far from equilibrium. Even if
equilibrium is reached in a heavy-ion collision this state changes
by cooling through expansion and particle emission and any observed
signal is built up by integrating over the emissions from all these
different stages of the reaction.

Calculations of in-medium properties of hadrons thus necessarily
rely on a number of simplifying assumptions, foremost being that of
an infinite medium at rest in which the hadron under study is
embedded. The properties so calculated are then, in a second step,
being locally inserted into a time-dependent event simulation.  In
actual experiments these hadrons are observed through their decay
products and these have to travel through the surrounding nuclear
matter to the detectors. Except for the case of electromagnetic
signals (photons, dileptons) this is connected with often sizeable
final state interactions (FSI) that have to be treated as realistic
as possible. For a long period the Glauber approximation which
allows only for absorptive processes along a straight-line path has
been the method of choice in theories of photonuclear reactions on
nuclei. This may be sufficient if one is only interested in total
yields of strongly absorbed particles. However, it is clearly
insufficient when one aims at, for example, reconstructing the
spectral function of a hadron inside matter through its decay
products. Rescattering and sidefeeding through coupled channel
effects can affect the final result so that a realistic description
of such effects is absolutely mandatory \cite{FalterShad}.

In this lecture note I first outline the fundamentals of chiral
symmetry, its expected density dependence and its connection with
actual observables. I then summarize results that we have obtained
in studies of observable consequences of in-medium changes of
hadronic spectral functions in reactions of elementary probes with
nuclei. I demonstrate that the expected in-medium sensitivity in
such reactions is as high as that in relativistic heavy-ion
collisions and that in particular photonuclear reactions present an
independent, cleaner testing ground for assumptions made in
analyzing heavy-ion reactions.

\section{Theory}

\subsection{Chiral Symmetry}

A large part of the current interest in in-medium properties of
hadrons comes from the hope to learn something about quarks in
nuclei. More specifically, one hopes to see precursors of a
restoration of the original symmetries of the theory of strong
interactions, i.e.\ QCD, which are spontaneously broken in our
world. In this section of the lecture I briefly summarize the -- for
the present discussion -- most relevant features of QCD. This
material here draws heavily on my book \cite{Moselbuch} where more
many details can be found.

The QCD Lagrangian
\be                                                              \label{LQCD}
{\cal L}_{\rm QCD} = - \frac{1}{4} G^c_{\mu \nu }{G^c}^{\mu \nu } +
\sum \limits_{f} \left[ \bar q_f ({\rm i} \gamma^\mu D_\mu -  m_f)
q_f \right]
     \quad
\ee
describes all the strong interactions of color-triplets $q$
representing the quarks. Here the sum over $c$ runs over the eight
members of the color octet and $D_\mu $ is given by
\be
D_\mu  = \partial _\mu  + {\rm i} g \frac{\lambda^c}{2} G^c_\mu
\quad .
\ee
The $G^c_\mu $ represent the eight vector fields describing the
gauge bosons of the theory, called ``gluons", because they transmit
the binding forces and $G_{\mu \nu}$ in (\ref{LQCD}) denotes the
corresponding field thensor. $g$ is a dimensionless coupling
constant (for further details see Chap.\ 15 in \cite{Moselbuch}).

If one assumes that the masses of the up- and down-quarks are nearly
the same, then QCD possesses a global SU(2)$_V$ symmetry under the
transformation
\be                                                       \label{SU3flav}
q_c(x) \rightarrow  q'_c(x) = \sum_{f=1}^3 {\rm e}^{-{\rm i}
\varepsilon^f \frac{\lambda ^f}{2}} q_c(x) \quad .
\ee
which acts onto the flavor degrees of freedom of the quark spinor
given by
\be
q_c(x) = \left( \begin{array}{c} u_c(x) \\
                                 d_c(x) \\
                                 s_c(x)  \end{array}
         \right)  \quad ,
\ee
where $u_c$, $d_c$ and $s_c$ represent the up, down and strange
quark, respectively, with a color index $c$. The sum over the flavor
index $f$ in equation (\ref{SU3flav}) runs only from 1 to 3, i.e.\
over the generators $\lambda ^f$ of the $SU(2)$ subgroup of the full
flavor-$SU(3)$. As a consequence of the invariance of ${\cal L}_{\rm
QCD}$ under the transformation (\ref{SU3flav}), the flavor vector
current
\be
V^f_\mu (x) =  \bar q_c \gamma _\mu  \frac{\lambda ^f}{2} q_c \quad
               (f = 1,2,3)
\ee
is conserved on the quark level. The observed isospin symmetry of
strong interactions and the corresponding vector current
conservation (CVC) can thus be understood as a consequence of a
symmetry of quark-quark interactions.

If, in addition, $m_u = m_d = 0$, then there is also a symmetry of
the QCD Lagrangian under
\be                                                       \label{SU3flavax}
q_c(x) \rightarrow  q'_c(x) = \sum_{f=1}^3 {\rm e}^{-{\rm i}
\gamma_5 \varepsilon^f \frac{\lambda ^f}{2}} q_c(x) \quad ,
\ee
so that the axial vector current
\be
A^f_\mu  = \bar q_c \gamma _\mu \gamma _5 \frac{\lambda ^f}{2} q_c
\quad (f=1,2,3)
\ee
is also conserved. In these considerations one uses the fact that
the gluon fields $G^c_\mu $ are pure color fields and are thus not
affected by the flavor generators $\lambda ^f$ or by the Dirac
algebra, e.g.\ by $\gamma _5$.

It is clear that the chiral symmetry cannot be a manifest one on the
hadronic level, since the hadrons are all massive. However, the
Goldstone mechanism of spontaneous symmetry breaking can reconcile
the nonzero masses of the hadrons with the observed (partial)
conservation of the axial current. The axial $SU(2)$ symmetry is
then assumed to be realized in the Goldstone mode in which the
Lagrangian, but not the state of the system, possesses the axial
$SU(2)$ symmetry. Therefore, it is obvious that in the real world
also the chiral symmetry must be spontaneously broken. An order
parameter for this symmetry breaking is given by the so-called
chiral condensate
\be
\langle|\bar{q} q | \rangle \ne 0~,
\ee
where the expectation value is taken in the vacuum state. The chiral
condensate and thus the ground state can be chirally symmetric only
if this condensate is zero as one can show by applying the
transformation (\ref{SU3flavax}) (see also discussion below).

A very simple estimate shows that the chiral condensate in the
nuclear medium is in lowest order in density given by \cite{Wambach}
\begin{equation}     \label{qbarq}
\langle \bar{q} q \rangle_{\rm med}(\rho,T) \approx \left( 1 -
\sum_h \frac{\Sigma_h \rho^s_h(\rho,T}{f_\pi^2 m_\pi^2} \right)
\langle \bar{q} q \rangle_{\rm vac} ~.
\end{equation}
Here $\rho_s$ is the \emph{scalar} density of the hadron $h$ in the
nuclear system and $\Sigma_h$ the so-called sigma-commutator that
contains information on the chiral properties of $h$. The sum runs
over all hadronic states. While (\ref{qbarq}) is nearly exact, its
actual value is limited because neither the sigma-commutators of the
higher lying hadrons nor their scalar densities are known. However,
at very low temperatures close to the groundstate these are
accessible for the nucleon. Here $\rho_s \approx \rho_v \frac{m}{E}$
so that the condensate drops linearly with the nuclear (vector)
density. Inserting numerical values for the physical constants in
(\ref{qbarq}) gives
\be
\langle \bar{q} q \rangle_{\rm med}(\rho,0) \approx \left( 1 - 0.3
\frac{\rho}{\rho_0} \right) \langle \bar{q} q \rangle_{\rm vac} ~.
\ee

This drop of the chiral condensate with density can be understood in
physical terms: with increasing density the hadrons with their
chirally symmetric phase in their interior fill in more and more
space in the vacuum with its spontaneously broken chiral symmetry.
Note that this is a pure volume effect; it is there already for a
free, non-interacting hadron gas. One can obtain a very simple
estimate for this effect by assuming that inside baryons the
chirally symmetric phase prevails, as it does in bag models (for a
review of such models see \cite{Moselbuch}). Assuming a baryon
radius of $\approx 0.8$ fm one obtains $V_N \approx 2.1$ fm$^3$.
This number has to be compared with the specific volume of one
baryon at normal nuclear matter density ($\rho_0 = 0.15\:
$fm$^{-3}$): $ v = 0.15^{-1}\: $fm$^{3}\approx 6.7\: $fm$^{3}$. This
shows that at about $v/V_N \approx 3$ times nuclear equilibrium
density the baryons significantly overlap so that their chirally
restored phase fills all space.

The so-called Nambu--Jona-Lasinio (NJL) model has also often been
used to determine the drop of the chiral condensate with density
(and temperature) and to establish a link between the chiral
condensate and physical masses. In this model the QCD Lagrangian is
replaced by a Lagrangian that exhibits the chiral symmetry of QCD
manifestly and in which the gluon fields are assumed to be
integrated out. The interactions mediated by them are modeled by  a
point-interaction of the quarks. The NJL model in the two-flavor
version is given by
\be                        \label{NJL}
{\cal L} = {\rm i} \bar{q} \gamma_\mu \partial^\mu q
           + G \left[ \left(\bar{q} q \right)^2
           + \left(\bar{q} {\rm i} \gamma_5 \vec{\tau} q \right)^2 \right]
           ~.
\ee
While the authors originally formulated this model in terms of
nucleon fields (quarks were not known then) we have given here the
Lagrangian in terms of quark fields $q$. The Lagrangian (\ref{NJL})
represents a system of massless fermions interacting through a
contact interaction (the term in square parentheses). The constant
$G$ is a coupling constant with the dimension of (mass)$^{-2}$; it
is assumed to be positive so that the self-interaction of the quark
fields is attractive. Since the interaction is local it corresponds
to a $\delta$-function two-body potential between the quarks in a
non-relativistic language.

In Hartree-Fock mean field theory the single particle Hamiltonian
can be obtained by a variation of the vacuum energy corresponding to
the NJL Lagrangian with respect to $\bar{q}$. The ensuing single
particle Hamiltonian exhibits a mass term
\be        \label{MNJL}
M q = \frac{\delta \langle{\cal H}\rangle}{\delta \bar{q}} = - 2 G
\langle \bar{q}q \rangle q ~.
\ee
The mass can be evaluated in a vacuum of free states
\bea                                           \label{Mgap}
M &=& \mbox{} -2 G \langle 0 | {\bar q} q | 0 \rangle     \nonumber \\
  &=& \mbox{} 2 G \frac{1}{V} \sum_{ps} \frac{M}{\sqrt{\vec{p}^2 + M^2}}   ~.
\eea
The sum over states in the Dirac sea diverges; it thus has to be
regularized by introducing an appropriate cut-off $\Lambda$ for the
sum over momenta. Going over to an integral representation gives
\be            \label{Meq}
M = 4 G  \int_{|\vec{p}| < \Lambda} \frac{M}{\sqrt{\vec{p}^2 + M^2}}
    \frac{d^3p}{(2 \pi)^3} ~.
\ee

It is obvious that (\ref{Meq}) has $M=0$ as a solution,
corresponding to a chirally symmetric vacuum (ground-) state.
However, for large enough couplings also a solution with $M > 0$ is
possible. In this case the vacuum state no longer exhibits the
chiral symmetry of the underlying theory, QCD; the symmetry is
spontaneously broken. The value of the 'chiral condensate'
(\ref{MNJL})
\be
\langle \bar{q}q \rangle = - \frac{1}{V} \sum_{ps}
\frac{M}{\sqrt{\vec{p}^2 + M^2}}
\ee
gives a measure for the amount of chiral symmetry breaking: it is
zero if chiral symmetry is present in the vacuum and nonzero
otherwise.
\begin{figure} \label{MNJLfig}
\epsfig{file=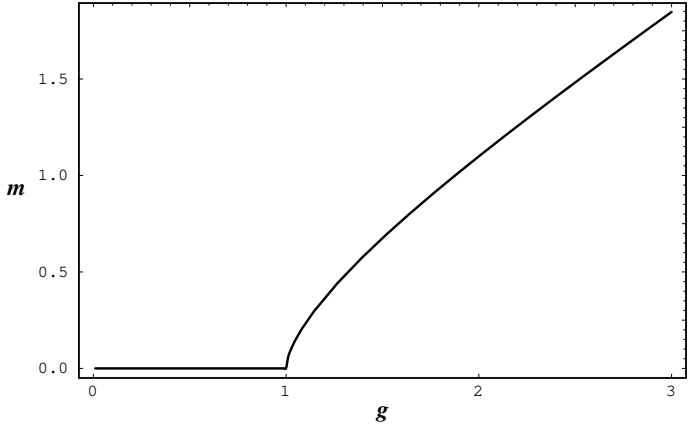,height=6cm} \caption{Scaled quark mass $m =
M/\Lambda$ in the NJL model as a function of the scaled coupling
constant $g = \Lambda^2 / \pi^2\, G$ for vanishing Fermi momentum
(from \protect\cite{Moselbuch})}
\end{figure}
Equation (\ref{Meq}) has exactly the same structure as the gap
equation in the BCS theory of superconductivity, used, for example,
to obtain the pairing gap as a function of the pairing interaction
strength. The mass $M$ plays here the role of the the pairing gap
$\Delta$.

It is a well-known property of the gap equation that the gap
decreases -- and ultimately disappears -- due to the blocking effect
when states above the Fermi-level become occupied, either through a
thermal excitation of the system or through the presence of odd
nucleons. Exactly the same phenomenon appears here: if also
positive-energy quark states are occupied, either through a
temperature in the system or through a non-vanishing baryon density,
the integral on the rhs of (\ref{Meq}) extends over these states as
well; because of the properties of the positive-energy spinors these
states appear with an opposite sign and give a contribution to the
mass
\be
\Delta M = - 4 G \int_{|\vec{p}| < p_{\rm F}}
                 \frac{M}{\sqrt{\vec{p}^2 + M^2}}
                 \frac{d^3p}{(2 \pi)^3}  ~.
\ee
In order to compensate for this negative contribution, with $G$
fixed, the masses then have to become smaller until they have to
vanish altogether, when the occupation of the positive energy states
is increased more and more. Thus, with increasing density the quarks
become massless again and chiral symmetry is restored. The same
effect takes place when the temperature of the system is increased.
In that case states above the Fermi momentum are partially occupied
and this blocking leads to a decrease of the mass, just as in BCS
theory.

\subsection{QCD Sum Rules and Hadronic Models}
In the NJL model the dropping of the chiral condensate with density
and/or temperature directly causes a drop of the mass because both
are linearly proportional. This is no longer the case in complex
hadronic systems. How the drop of the scalar condensate there
translates into observable hadron masses is not uniquely prescribed.
The only rigorous connection is given by the QCD sum rules that
relates an integral over the hadronic spectral function to a sum
over combinations of quark- and gluon-condensates with powers of
$1/Q^2$. The starting point for this connection is the
electromagnetic current-current correlator
\begin{equation}
\Pi_{\mu\nu}(q) = i \int\!\! d^4\!x \, e^{iqx} \langle T j_\mu(x)
j_\nu(0) \rangle    \, ,
  \label{eq:curcur}
\end{equation}
where the current is expressed in terms of quark-field operators.
This correlator can be decomposed \cite{Leupold}
\begin{equation}
  \label{eq:decompmunu}
\Pi_{\mu\nu}(q) = q_\mu q_\nu R(q^2) - g_{\mu\nu} \Pi^{\rm
isotr}(q^2)    \,.
\end{equation}
The QCD sum rule then reads
\begin{eqnarray}
R^{{\rm OPE}}(Q^2) & = & {\tilde c_1 \over Q^2} + \tilde c_2 -{Q^2
\over \pi} \int\limits_0^\infty \!\! ds \, {{\rm Im}R^{{\rm HAD}}(s)
\over (s+Q^2)s}
  \label{eq:opehadr}
\end{eqnarray}
with $Q^2:= -q^2 \gg 0$ and some subtraction constants $\tilde c_i$.
Here $R^{\rm OPE}$ represents a Wilson's operator expansion (OPE) of
the current-current correlator in terms of quark and gluon degrees
of freedom in the space-like region. On the other hand, $R^{{\rm
HAD}}(s)$ in (\ref{eq:opehadr}) is the same object for time-like
momenta, represented by a parametrization in terms of hadronic
variables. Experimentally, it can be determined by hadron production
in $e^+e^-$ reactions \cite{Peskin-Schroeder}. It is usually written
in a form that exhibits the asymptotic, high momentum behavior of
QCD in a manifest form:
\begin{eqnarray}
{\rm Im}R^{{\rm HAD}}(s) & = & \Theta(s_0 -s) \, {\rm Im}R^{{\rm
RES}}(s) + \Theta(s -s_0) \, {1 \over 8 \pi} \left( 1+ {\alpha_s
\over \pi} \right) ~,
  \label{eq:pihadans}
\end{eqnarray}
where $s_0$ denotes the threshold between the low energy region
described by a spectral function for the lowest lying resonance,
${\rm Im}R^{{\rm RES}}$, and the high energy region described by a
continuum calculated from perturbative QCD. The second term on the
rhs of (\ref{eq:pihadans}) represents the QCD perturbative result
that survives when $s \to \infty$.

The dispersion integral in (\ref{eq:opehadr}) connects time- and
space-like momenta. Eq. (\ref{eq:opehadr}) also connects the
hadronic (rhs) with the quark (lhs) world. It allows -- after the
technical step of a Borel transformation -- to determine parameters
in a hadronic parametrization of $R^{{\rm HAD}}(s)$ by comparing the
lhs of this equation with its rhs.

The operator product expansion of $R^{\rm OPE}$ on the lhs involves
a separation of momentum scales between the hard scale $Q^2$ and the
soft, nonperturbative quark- and gluon condensates
\cite{Leupold,LeupoldMosel}. Using the measured, known vacuum
spectral function for $R^{\rm HAD}$ allows one to obtain information
about the condensates appearing on the lhs of (\ref{eq:opehadr}).

Turning this argument around one can model the density dependence of
the quark condensates on the lhs of (\ref{eq:opehadr}) by using, for
example, the density dependence obtained in the NJL model.  The
QCDSR then gives information on the density dependence of the
hadronic spectral function $R^{\rm HAD}$ on the rhs of
(\ref{eq:opehadr}).

Since the spectral function appears under an integral the
information obtained is, however, not very specific. However,
Leupold et al. have shown \cite{Leupold,LeupoldMosel} that the QCDSR
provides important constraints for the hadronic spectral functions
in medium, but it does not fix them. Recently Kaempfer et al have
turned this argument around by pointing out that measuring an
in-medium spectral function of the $\omega$ meson could help to
determine the density dependence of the chiral condensate
\cite{Kaempfer}.

Thus models are needed for the hadronic interactions. The
quantitatively reliable ones can at present be based only on
'classical' hadrons and their interactions. Indeed, in lowest order
in the density the mass and width of an interacting hadron in
nuclear matter at zero temperature and vector density $\rho_v$ are
given by (for a meson, for example)
\begin{eqnarray}     \label{trho}
{m^*}^2 = m^2 - 4 \pi \Re f_{m N}(q_0,\theta = 0)\, \rho_v
\nonumber \\
m^* \Gamma^* = m \Gamma^0 -  4 \pi \Im f_{mN}(q_0,\theta = 0)\,
\rho_v ~.
\end{eqnarray}
Here $f_{mN}(q_0,\theta = 0)$ is the forward scattering amplitude
for a meson with energy $q_0$ on a nucleon. The width $\Gamma^0$
denotes the free decay width of the particle. For the imaginary part
this is nothing other than the classical relation $\Gamma^* -
\Gamma^0 = v \sigma \rho_v$ for the collision width, where $\sigma$
is the total cross section. This can easily be seen by using the
optical theorem.

Actually evaluating mass and width from (\ref{trho}) requires
knowledge of the scattering amplitude which can only be obtained
from very detailed analyses of experiments. The $s$-channel
contributions to this scattering amplitude are determined by the
properties of nucleon resonances and these are often not very well
known yet. Here, resonance physics meets in-medium physics.

Unfortunately it is not a-priori known up to which densities the
low-density expansion (\ref{trho}) is useful. Post et al.
\cite{Postneu} have recently investigated this question in a
coupled-channel calculation of selfenergies. Their analysis
comprises pions, $\eta$-mesons and $\rho$-mesons as well as all
baryon resonances with a sizeable coupling to any of these mesons.
The authors of \cite{Postneu} find that already for densities less
than $0.5 \rho_0$ the linear scaling of the selfenergies inherent
in (\ref{trho}) is badly violated for the $\rho$ and the $\pi$
mesons, whereas it is a reasonable approximation for the $\eta$
meson. Reasons for this deviation from linearity are Fermi-motion,
Pauli-blocking, selfconsistency and short-range correlations. For
different mesons different sources of the discrepancy prevail: for
the $\rho$ and $\eta$ mesons the iterations act against the
low-density theorem by inducing a strong broadening for the
$D_{13}(1520)$ and a slightly repulsive mass shift for the
$S_{11}(1535)$ nucleon resonances to which the $\rho$ and the
$\eta$ meson, respectively, couple. The investigation of in-medium
properties of mesons, for example, thus involves at the same time
the study of in-medium properties of nucleon resonances and is
thus a coupled-channel problem.

Note that such a picture, in which the selfenergies of hadrons are
generated by interactions with the surrounding baryons, also
encompasses the change of the chiral condensate in (\ref{qbarq}),
obtained there for non-interacting hadrons. If the spectral function
of a non-interacting hadron changes as a function of density, then
in a classical hadronic theory, which works with fixed (free) hadron
masses, this change will show up as an energy-dependent interaction
and is thus contained in any empirical phenomenological cross
section.

\subsection{Coupled Channel Treatment of Incoherent Particle
Production}\label{sec:CoupledChannel}

Very high nuclear densities ($2 - 8\rho_0$) and temperatures $T$
up to or even higher than $\approx$ 170 MeV can be reached with
present day's accelerators in heavy-ion collisions. Thus, any
density-dependent effect gets magnified in such collisions.
However, the observed signal always represents a time-integral
over various quite different stages of the collision --
non-equilibrium and equilibrium, the latter at various densities
and temperatures. The observables thus have to be modelled in a
dynamic theory. In contrast, the theoretical input is always
calculated under the simplifying assumption of a hadron in
stationary nuclear matter in equilibrium and at fixed density. The
results of such calculations are then used in dynamical
simulations of various degrees of sophistication most of which
invoke a quasi-stationary approximation. In order to avoid these
intrinsic difficulties we have looked for possible effects in
reactions that proceed closer to equilibrium, i.e. reactions of
elementary probes such as protons, pions, and photons on nuclei.
The densities probed in such reactions are always $\le \rho_0$,
with most of the nucleons actually being at about $0.5 \rho_0$. On
the other hand, the target is stationary and the reaction proceeds
much closer to (cold) equilibrium than in a relativistic heavy-ion
collision. If any observable effects of in-medium changes of
hadronic properties survive, even though the densities probed are
always $\le \rho_0$, then the study of hadronic in-medium
properties in reactions with elementary probes on nuclei provides
an essential baseline for in-medium effects in hot nuclear matter
probed in ultra-relativistic heavy-ion collisions.

With the aim of exploring this possibility we have over the last few
years undertaken a number of calculations for proton-
\cite{Bratprot}, pion- \cite{Weidmann,Effepi} and photon-
\cite{Effephot} induced reactions. All of them have one feature in
common: they treat the final state incoherently in a coupled channel
transport calculation that allows for elastic and inelastic
scattering of, particle production by and absorption of the produced
hadrons. A new feature of these calculations is that hadrons with
their correct spectral functions can actually be produced and
transported consistently. This is quite an advantage over earlier
treatments \cite{Brat-Cass,Fuchs} in which the mesons were always
produced and transported with their pole mass and their spectral
function was later on folded in only for their decay. The method is
summarized in the following section, more details can be found in
\cite{Effephot}.

We separate the photonuclear reaction into three steps. First, we
determine the amount of shadowing for the incoming photon; this
obviously depends on its momentum transfer $Q^2$. Second, the
primary particle is produced and third, the produced particles are
propagated through the nuclear medium until they leave the
nucleus.

\paragraph{Shadowing.} Photonuclear reactions show shadowing in the
entrance channel, for real photons from an energy of about 1 GeV
on upwards \cite{Bianchi}. This shadowing is due to a coherent
superposition of bare photon and vector meson components in the
incoming photon and is handled here by means of a Glauber multiple
scattering model \cite{FalterShad}. In this way we obtain for each
value of virtuality $Q^2$ and energy $\nu$ of the photon a spatial
distribution for the probability that the incoming photon reaches
a given point; for details see \cite{FalterShad,Effe,Falterinc}.

\paragraph{Initial Production.}
The initial particle production is handled differently depending
on the invariant mass $W = \sqrt{s}$ of the excited state of the
nucleon. If $W < 2$ GeV, we invoke a nucleon resonance model that
has been adjusted to nuclear data on resonance-driven particle
production \cite{Effephot}. If $W > 2$ GeV the particle yield is
calculated with standard codes developed for high energy nuclear
reactions, i.e.\ FRITIOF or PYTHIA; details are given in
\cite{Falterneu}. We have made efforts to ensure a smooth
transition of cross sections in the transition from resonance
physics to DIS.

\paragraph{Final State Interactions.}
The final state is described by a semiclassical coupled channel
transport model that had originally been developed for the
description of heavy-ion collisions and has since then been
applied to various more elementary reactions on nuclei with
protons, pions and photons in the entrance channel.

In this method the spectral phase space distributions of all
particles involved are propagated in time, from the initial first
contact of the photon with the nucleus all the way to the final
hadrons leaving the nuclear volume on their way to the detector.
The spectral phase space distributions
$F_h(\vec{x},\vec{p},\mu,t)$ give at each moment of time and for
each particle class $h$ the probability to find a particle of that
class with a (possibly off-shell) mass $\mu$ and momentum
$\vec{p}$ at position $\vec{x}$. Its time-development is determined
by the BUU equation
\be     \label{BUU}
(\frac{\partial}{\partial t} + \frac{\partial H_h}{\partial
\vec{p}} \frac{\partial}{\partial \vec{r}} - \frac{\partial
H_h}{\partial \vec{r}} \frac{\partial}{\partial \vec{p}})F_h=G_h
{\cal A}_h - L_h F_h.
\ee
Here $H_h$ gives the energy of the hadron $h$ that is being
transported; it contains the mass, the selfenergy (mean field) of
the particle and a term that drives an off-shell particle back to
its mass shell. The terms on the lhs of (\ref{BUU}) are the
so-called \emph{drift terms} since they describe the independent
transport of each hadron class $h$. The terms on the rhs of
(\ref{BUU}) are the \emph{collision terms}; they describe both
elastic and inelastic collisions between the hadrons. Here the
term \emph{inelastic collisions} includes those collisions that
either lead to particle production or particle absorption. The
former is described by the \emph{gain term} $G_h {\cal A}_h$ on
the rhs in (\ref{BUU}), the latter process (absorption) by the
\emph{loss term} $L_h F_h$. Note that the gain term is
proportional to the spectral function $\mathcal{A}$ of the
particle being produced, thus allowing for production of off-shell
particles. On the contrary, the loss term is proportional to the
spectral phase space distribution itself: the more particles there
are the more can be absorbed. The terms $G_h$ and $L_h$ on the rhs
give the actual strength of the gain and loss terms, respectively.
They have the form of Born-approximation collision integrals and
take the Pauli-principle into account. The free collision rates
themselves are taken from experiment or are calculated
\cite{Effephot}.

Eq.\ (\ref{BUU}) contains a selfconsistency problem. The collision
rates embedded in $G$ and $L$ determine the collisional broadening
of the particles involved and thus their spectral function
$\mathcal{A}$. The widths of the particles, resonances or mesons,
thus evolve in time away from their vacuum values. In addition,
broad particles can be produced off their peak mass and then
propagated. The extra 'potential' in $H$ already mentioned ensures
that all particles are being driven back to their mass-shell when
they leave the nucleus. The actual method used is described in
\cite{Effephot}. It is based on an analysis of the Kadanoff-Baym
equation that has led to practical schemes for the propagation of
off-shell particles \cite{Leupoldoffshell,JuchemCassing}´. The
possibility to transport off-shell particles represents a major
breakthrough in this field. For further details of the model see
Ref. \cite{Effephot} and \cite{Falterneu} and references therein.

\section{Particle Production on Nuclei -- Observables}

\section{\it $\eta$ Production} We first look at the prospects of
using reactions with hadronic final states and discuss the
photo-production of $\eta$-mesons on nuclei as a first example.
These mesons are unique in that they are sensitive to one
dominating resonance the $S_{11}(1535)$ so that one may hope to
learn something about the properties of this resonance inside
nuclei. Experiments for this reaction were performed both by the
TAPS collaboration \cite{TAPS,Krusche} and at KEK \cite{KEK}.

Estimates of the collisional broadening of the $S_{11}(1535)$
resonance have given a collision width of about 35 MeV at $\rho_0$
\cite{EffeHom}. The more recent, and more refined, selfconsistent
calculations of \cite{Postneu} give a very similar value for this
resonance. In addition, a dispersive calculation of the real part
of the selfenergy for the resonance at rest gives only an
insignificant shift of the resonance position. Thus any
momentum-dependence of the selfenergy as observed in
photon-nucleus data can directly be attributed to binding energy
effects \cite{Lehreta}. The results obtained in \cite{Lehreta}
indicate that the momentum dependence of the $N^*(1535)$ potential
has to be very similar to that of the nucleons.

\begin{figure}[ht]

\centerline{\epsfig{file=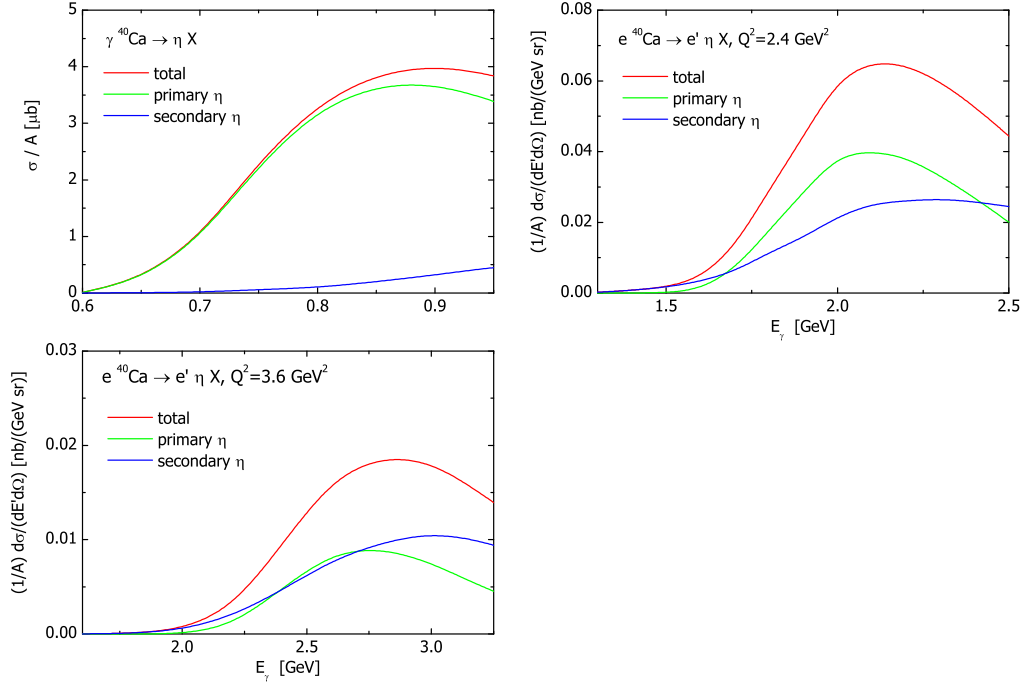,scale=1.2}}
\caption{Eta-electroproduction cross sections on $^{40}Ca$ for three
values of $Q^2$ given in the figures. The uppermost curve in all
three figures gives the total observed yield, the next curve from
top gives the contribution of the direct $\eta$ production channel
and the lowest curve shows the contribution of the secondary process
$\pi N \rightarrow \eta N$. For the highest value of $Q^2=3.6$
GeV$^2$ the primary and secondary production channels nearly
coincide (from \protect\cite{Lehrelectro}).}\label{fig:electroeta}

\end{figure}

The particular advantage of our coupled-channels approach can be
seen in Fig.\ \ref{fig:electroeta} which shows results both for
photon- and electroproduction of $\eta$'s on nuclei; for the
latter process so far no data are available.
The calculations give the interesting result that for
photo-production a secondary reaction channel becomes important at
high energies and virtualities. In this channel first a pion is
produced which travels through the nucleus and through $\pi N
\rightarrow N^*(1535) \rightarrow N \eta$ produces the $\eta$ that
is finally observed in experiment. This channel becomes even
dominant for high $Q^2$ electroproduction. In both cases the
reason for the observed growing of the importance of secondary
production channels lies in the higher momentum transferred to the
initial pion \cite{Lehrelectro}.

\subsection{$2\pi$ Production}
If chiral symmetry is restored, the masses of the scalar
isoscalar $\sigma$ meson and that of the scalar isovector pion
should become degenerate. This implies that the spectral function
of the $\sigma$ should become softer and narrower with its
strength moving down to the $2\pi$ threshold. This leads to a threshold
enhancement in the $\pi\pi$ invariant mass spectrum due to suppression
of the phase-space for the $\sigma\rightarrow\pi\pi$ decay.

A first measurement of the two pion invariant mass spectrum has been
obtained by the CHAOS collaboration in pion induced reactions on
nuclei \cite{CHAOS}. The authors of \cite{CHAOS} claimed to indeed have
seen an accumulation of spectral strength near the $2\pi$ threshold for
heavy target nuclei in the $\pi^+\pi^-$ mass distribution. According to the
arguments presented in the introduction, photon induced reactions in nuclei
are much better suited to investigate double pion pion production at finite
baryon densities. A recent
experiment with the TAPS spectrometer at the tagged-photon facility MAMI-B in
Mainz indeed shows an even more pronounced accumulation of spectral strength of the two
pion mass spectrum for low invariant masses with increasing target
mass corresponding to increasing average densities probed
\cite{TAPSsigma}. This accumulation has been observed for the
$\pi^0\pi^0$ but not for the $\pi^\pm \pi^0$ final state.

These results have been explained by a model developed by Roca et al.
\cite{Oset2pi}.  In this model the $\sigma$ meson is generated dynamically
as a resonance in the $\pi\pi$ scattering amplitude. By dressing the pion
propagators in the medium by particle-hole loops, they found the observed
downward shift of the $\pi\pi$ mass spectrum to be consistent with a dropping of
the $\sigma$-pole in the $\pi\pi$ scattering amplitude, i.~e. a lowering of
the $\sigma$ meson mass.

\begin{figure}[h]

\centerline{\epsfig{file=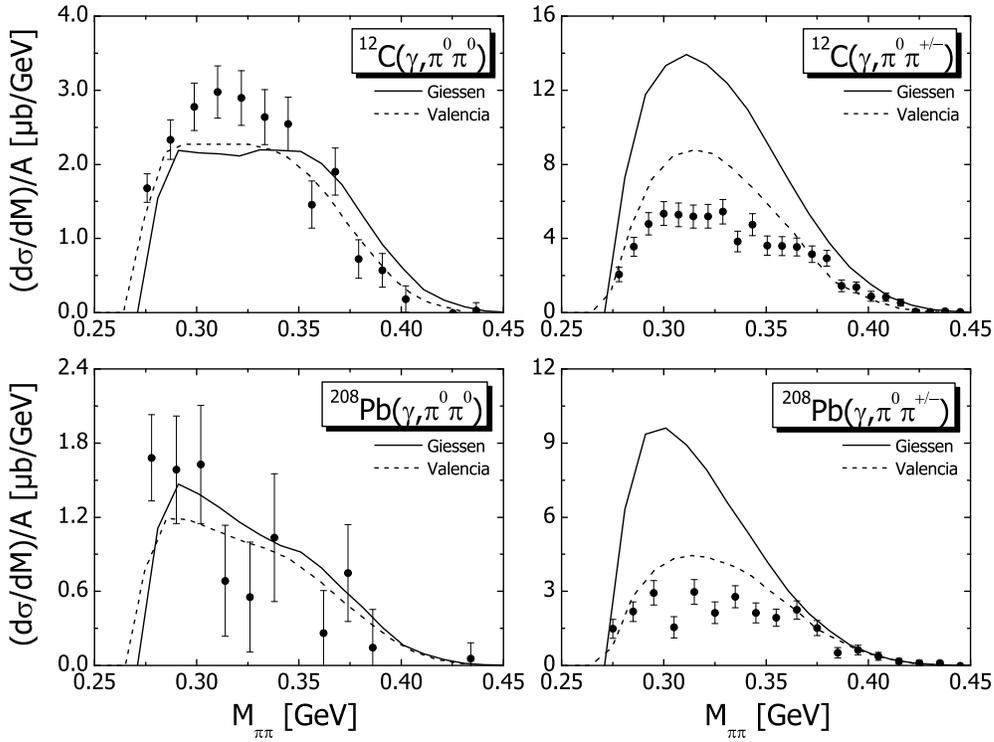,scale=0.9}}

\caption{Two pion invariant mass distributions for the $\pi^0\pi^0$
(left) and $\pi^0\pi^\pm$ (right) photoproduction on $^{12}C$ and
$^{208}Pb$ (from \protect\cite{Ruso}). The solid lines represent our
results whereas the dashed lines labelled ``Valencia'' depict the
results of ref. \protect\cite{Oset2pi}.}
 \label{fig:2pi}

\end{figure}

In analyzing the TAPS data it is absolutely essential to simulate
the final state interactions of the outgoing pions, correlated or
not, in an as realistic way as possible. We have, therefore, taken
the conservative approach of analyzing this reaction without any
changes of the $\sigma$ spectral function in the
outgoing channel. The result of this study \cite{Ruso} is shown in
Fig.~\ref{fig:2pi}.

Fig.~\ref{fig:2pi} shows on its left side that the observed
downward shift of the $2\pi^0$ mass spectrum can be well
reproduced by final state interactions on the independent pions
without any in-medium modification of the $\pi\pi$ interaction.
This shift can be attributed to a
slowing down of the pions due to quasi elastic collisions with the
nucleons in the surrounding nuclear medium. Also shown is the result of the
calculations of Roca et al. \cite{Oset2pi}. Both
calculations obviously agree with each other so that the final
observables are fairly independent of any $\pi\pi$ correlations.

On the right side of Fig.~\ref{fig:2pi} the results for the
semi-charged $\pi^0\pi^\pm$ channel are shown. Here our theoretical
result (solid line) overestimates the data by up to a factor of 3.
The result obtained in Ref.~\cite{Oset2pi} also lies too high, but
by a lesser amount; the difference between the two calculations has
been explained in \cite{Ruso} by the neglect of charge-transfer
channels in the calculations of Roca et al. \cite{Oset2pi}. The
observed discrepancy with experiment for this channel is astounding
since the method used normally describes data within a much narrower
error band. Thus understanding this discrepancy is absolutely
essential before a shift or non-shift in the mass distribution for
this channel can be ascertained. A new analysis of the data may be
helpful in this regard \cite{Susan}.

Even before this problem has been resolved it is clear now that pion
rescattering, including charge transfer, plays a major role for the
$2\pi$ mass distributions. It is, therefore, absolutely  mandatory
that any theoretical evaluation of this reaction takes this FSI
effect into account. Because the reaction is incoherent quantum
mechanical approaches to this problem are not feasible. The
semiclassical transport approaches, on the other hand, suffer from
intrinsic problems at very low pion energies (in the present
experiment $\approx 10$ MeV kinetic energy). Here, first, the {\it
de Broglie} wavelength of the pion is large so that semiclassical
approaches have to be critically examined. Second, for these low
energies the low-energy real pion potential becomes important and
has to be included together with the absorptive part. Careful
analyses and comparisons with optical model approaches have shown
\cite{Buss} that both effects introduce some systematic error into
the calculations that have to be taken into account in drawing any
conclusions from this experiment.

\subsection{\it $\omega$ Production}

Many of the early studies of hadronic properties in medium
concentrated on the $\rho$-meson \cite{Peters,Klingl}, partly
because of its possible significance for an interpretation of the
CERES experiment. It is clear by now, however, that the dominant
effect on the in-medium properties of the $\rho$-meson is
collisional broadening that overshadows any possible mass shifts
\cite{Postneu} and is thus experimentally hard to observe. The
emphasis has, therefore, shifted to the $\omega$ meson. An
experiment measuring the $A(\gamma,\omega \rightarrow
\pi^0\gamma')X$ reaction is presently being analyzed by the
TAPS/Crystal Barrel collaborations at ELSA \cite{Messch}. The
varying theoretical predictions for the $\omega$ mass (640-765
MeV) \cite{Klingl} and width (up to 50 MeV) \cite{Weidmann,Friman}
in nuclear matter at rest encourage the use of such an exclusive
probe to learn about the $\omega$ spectral distribution in nuclei.

\begin{figure}[h]

\centerline{\epsfig{file=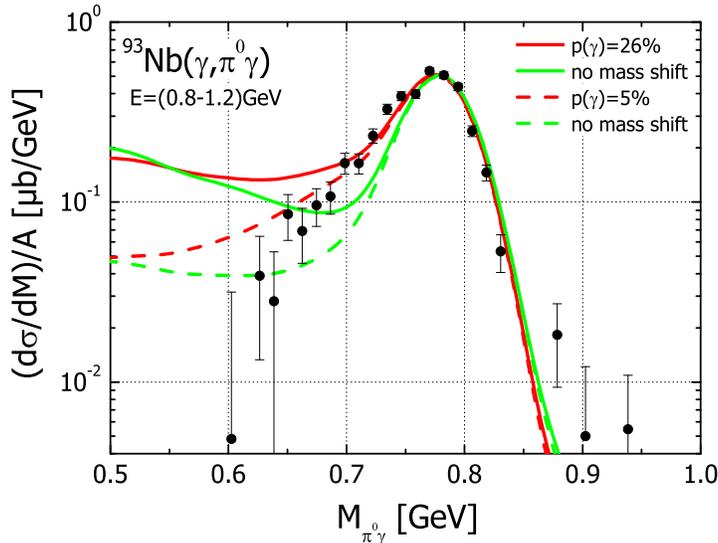,scale=1.0}}

\caption{Mass differential cross section for $\pi^0\gamma$
photoproduction off $^{93}$Nb. Shown are results both with and
without a mass-shift as explained in \protect\cite{MuehlOm}. The
quantity $p$ gives the escape probability for one of the four
photons in the $2\pi^0$ channel. The two solid curves give results
of calculations with p = 26 \% with and without mass-shift, the two
dashed curves give the same for p = 5\% (from
\protect\cite{Muehlichpriv}\label{omega}}
\end{figure}

Simulations have been performed at 1.2 GeV and 2.5 GeV photon
energy, which cover the accessible energies of the TAPS/Crystal
Barrel experiment. After reducing the combinatorial and rescattering
background by applying kinematic cuts on the outgoing particles, we
have obtained rather clear observable signals for an assumed
dropping of the $\omega$ mass inside nuclei \cite{MuehlOm}.
Therefore, in this case it should be possible to disentangle the
collisional broadening from a dropping mass.

One of the problems in the experimental analysis is the proper
subtraction of the background observed \cite{Trnka}; this has so far
been done in a heuristic way \cite{TrnkaPRL}. Our calculations
represent complete event simulations. It is, therefore, possible to
calculate these background contributions and to take experimental
acceptance effects into account. An example is shown in
Fig.~\ref{omega} which shows the effects of a possible
misidentification of the $\omega$ meson. This misidentification can
come about through the $2 \pi^0 \rightarrow 4 \gamma$ channel if one
of the four photons escapes detection and the remaining three
photons are identified as stemming from the $\pi^0 \gamma
\rightarrow 3 \gamma$ decay channel of the $\omega$-meson. The
calculations show that the misidentification does not affect the
low-mass side of the omega spectral function.

Fig.\ \ref{omega} shows a good agreement between the data of the
TAPS/CB@ELSA collaboration \cite{TrnkaPRL} for a photon escape
probability of 5 \% and a mass shift $m_\omega = m_\omega^0 - 0.18
\,\rho/\rho_0$. In \cite{MuehlOm} we have also discussed the
momentum-dependence of the $\omega$-selfenergy in medium and have
pointed out that this could be accessible through measurements which
gate on different three-momenta of the $\omega$ decay products. The
data presented by D. Trnka at this workshop \cite{Trnka} confirm
this expectation.

\subsection{\it Dilepton Production}

Dileptons, i.e.\ electron-positron pairs, in the outgoing channel
are an ideal probe for in-medium properties of hadrons since they
experience no strong final state interaction. A first experiment
to look for these dileptons in heavy-ion reactions was the DLS
experiment at the BEVALAC in Berkeley \cite{DLS}. Later on, and in
a higher energy regime, the CERES experiment has received a lot of
attention for its observation of an excess of dileptons with
invariant masses below those of the lightest vector mesons
\cite{CERES}.
\begin{figure}[h]
\begin{center}
\begin{minipage}[t]{8 cm}
\centerline{\epsfig{file=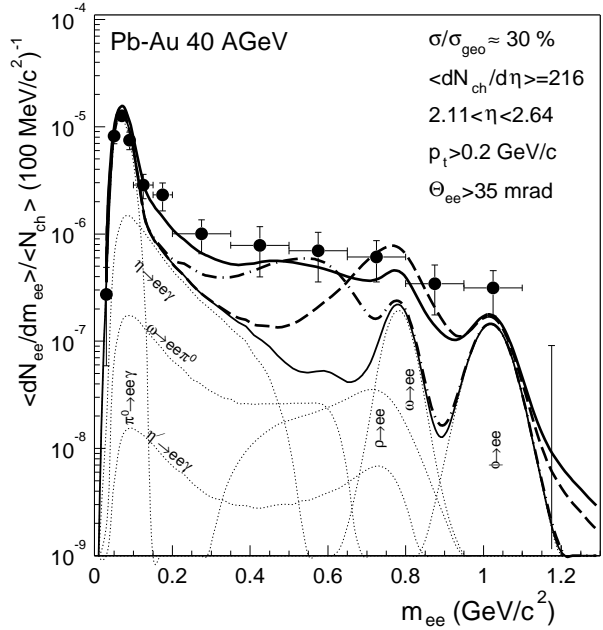,scale=0.8}}
\end{minipage}
\begin{minipage}[t]{16.5 cm}
\caption{Invariant  dilepton mass spectrum
 obtained with the CERES experiment in Pb + Au collisions at 40
 AGeV (from \protect\cite{CERES}). The thin curves give the contributions
 of individual hadronic sources to the total dilepton yield, the
 fat solid (modified spectral function) and the dash-dotted
 (dropping mass only) curves give the results of calculations
 \protect\cite{Rapp} employing an in-medium modified spectral function of the vector
 mesons.} \label{CERES}
\end{minipage}
\end{center}
\end{figure}
Explanations of this excess have focused on a change of in-medium
properties of these vector mesons in dense nuclear matter (see
e.g.\ \cite{Cassingdil,RappWam}). The radiating sources can be
nicely seen in Fig.~\ref{CERES} that shows the dilepton spectrum
obtained in a low-energy run at 40 AGeV together with the
elementary sources of dilepton radiation.

The figure exhibits clearly the rather strong contributions of the
vector mesons -- both direct and through their Dalitz decay -- at
invariant masses above about 500 MeV. The strong amplification of
the dilepton rate at small invariant masses $M$ caused by the
photon propagator, which contributes $\sim 1/M^4$ to the cross
section, leads to a strong sensitivity to changes of the spectral
function at small masses. Therefore, the excess observed in the
CERES experiment can be explained by such changes as has been
shown by various authors (see e.g.\ \cite{Brat-Cass} for a review
of such calculations).

In view of the uncertainties in interpreting these results
discussed earlier we have studied the dilepton photo-production in
reactions on nuclear targets. Looking for in-medium changes in
such a reaction is not \emph{a priori} hopeless: Even in
relativistic heavy-ion reactions only about 1/2 of all dileptons
come from densities larger than $2 \rho_0$ \cite{Brat-Cass}. In
these reactions the pion-density gets quite large in the late
stages of the collision. Correspondingly many $\rho$ mesons are
formed (through $\pi + \pi \to \rho$) late in the collision, where
the baryonic matter expands and its density becomes low again.

\begin{figure}
 \centerline{\epsfig{file=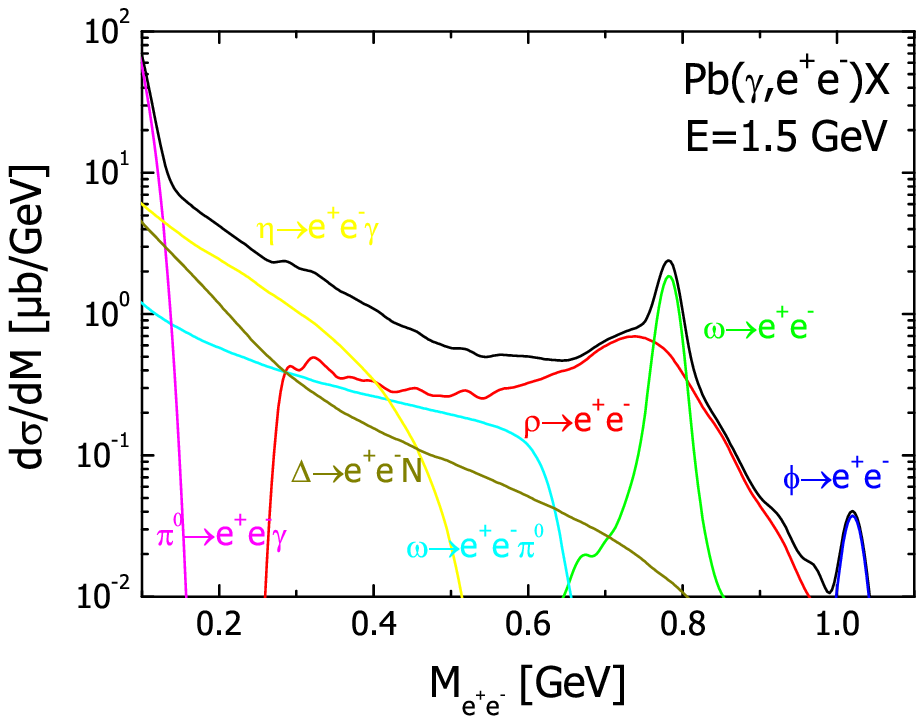,scale=1.0}}
\centerline{\epsfig{file=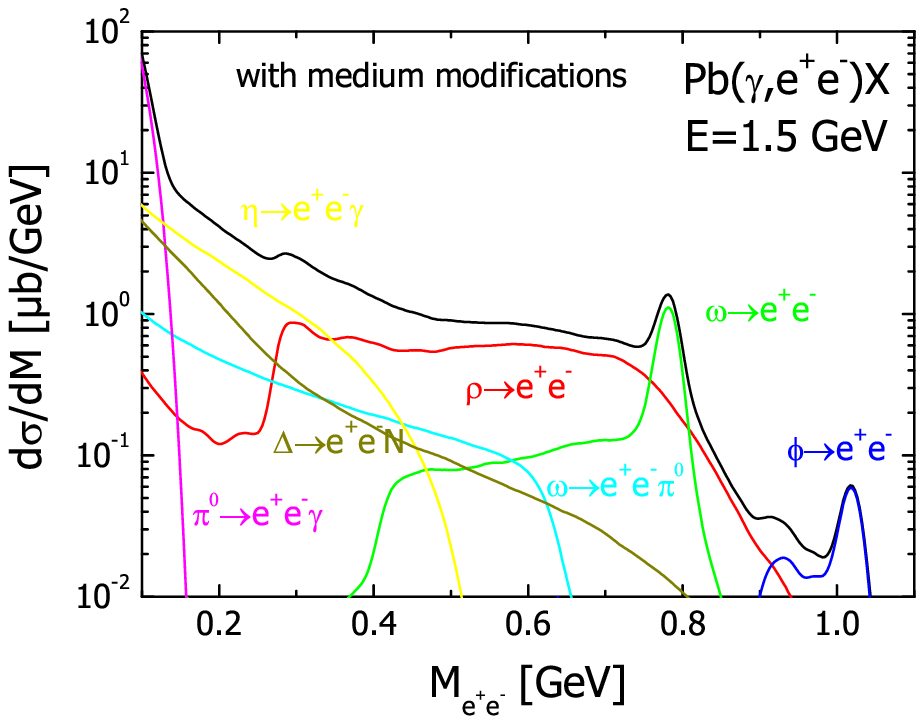,scale=1.0}}
\caption{Hadronic contributions to dilepton invariant mass spectra
for $\gamma + ^{208}Pb$ at 1.5 GeV photon energy). Indicated are the
individual contributions to the total yield; compare with
Fig.~\ref{CERES} (from \protect\cite{Muehlichpriv}).}
\label{Fige+e-}
\end{figure}
In \cite{Effephot} we have analyzed the photoproduction of
dileptons on nuclei in great detail. After removing the
Bethe-Heitler contribution the dilepton mass spectrum in a 2 GeV
photon-induced reaction looks very similar to that obtained in an
ultrarelativistic heavy-ion collision (Fig.\ \ref{CERES}). The
radiation sources are all the same in both otherwise quite
different reactions. The photon-induced reaction can thus  be used
as a baseline experiment that allows one to check crucial input
into the simulations of more complicated heavy-ion collision.

A typical result of such a calculation for the dilepton yield --
after removing the Bethe-Heitler component -- is given in Fig.\
\ref{Fige+e-}. The lower part of Fig.\ \ref{Fige+e-} shows that we
can expect observable effect of possible in-medium changes of the
vector meson spectral functions in medium on the low-mass side of
the $\omega$ peak. In \cite{Effephot} we have shown that these
effects can be drastically enhanced if proper kinematic cuts are
introduced that tend to enhance the in-medium decay of the vector
mesons. There it was shown that in the heavy nucleus $Pb$ the
$\omega$-peak completely disappears from the spectrum if in-medium
changes of width and mass are taken into account. The sensitivity of
such reactions is thus as large as that observed in
ultrarelativistic heavy-ion reactions.

An experimental verification of this prediction would be a major
step forward in our understanding of in-medium changes. The ongoing
g7 experiment at JLAB is presently analyzing such data
\cite{Weygand}. This experiment can also yield important information
on the time-like electromagnetic formfactor of the proton and its
resonances \cite{MoselHirsch95} on which little or nothing is known.

\section{Conclusions}\label{concl}
In this lecture note I have first outlined the theoretical
motivation for studies of in-medium properties of hadrons and their
relation to QCD. I have then shown that photonuclear reactions on
nuclei can give observable consequences of in-medium changes of
hadrons that are as big as those expected in heavy-ion collisions
which reach much higher energies, but proceed farther away from
equilibrium. Information from photonuclear reactions is important
and relevant for an understanding of high density -- high
temperature phenomena in ultrarelativistic heavy-ion collisions.
Special emphasis was put in these lectures not so much on the
theoretical calculations of hadronic in-medium properties under
simplified conditions, but more on the final, observable effects of
any such properties. I have discussed that for reliable predictions
of observables one has to take the final state interactions with all
their complications in a coupled channel calculation into account;
simple Glauber-type descriptions are not sufficient.

As an example that is free from complications by FSI I have shown
that in photonuclear reactions in the 1 - 2 GeV range the expected
sensitivity of dilepton spectra to changes of the $\rho$- and
$\omega$ meson properties in medium is as large as that in
ultrarelativistic heavy-ion collisions and that exactly the same
sources contribute to the dilepton yield in both experiments. While
the dilepton decay channel is free from hadronic final state
interactions this is not so when the signal has to be reconstructed
from hadrons present in the final state. While the $\omega$
photoproduction, identified by the semi-hadronic $\pi^0 \gamma$
decay channel, seems to exhibit a rather clean in-medium signal, the
double-pion experiments set up to look for an in-medium shift of
scalar strength in nuclei are dominated by final state interactions
of the produced pions that have to be properly taken into account in
any reliable analysis.

\section*{Acknowledgement}

I gratefully acknowledge many stimulating discussions with L.
Alvarez-Ruso, O. Buss, J. Lehr, S. Leupold, P. Muehlich and M. Post;
many of the results discussed in these lectures are based on their
work. This work has been supported by the Deutsche
Forschungsgemeinschaft, partly through the SFB/Transregio 16
``Subnuclear Structure of Matter''.

\end{document}